\documentclass[aps]{revtex4}
\usepackage{graphicx}% Include figure files
\usepackage{dcolumn}% Align table columns on decimal point
\usepackage{bm}% bold math
\usepackage{rotate}
\usepackage{epsfig}
\newcommand \be  {\begin{equation}}
\newcommand \bea {\begin{eqnarray} \nonumber }
\newcommand \ee  {\end{equation}}
\newcommand \eea {\end{eqnarray}}
\begin{document}
\title{Theory of collective opinion shifts: from smooth trends to abrupt
swings}
\author{Quentin Michard$^\dagger$}
\author{Jean-Philippe Bouchaud$^{+}$}
\email{bouchau@spec.saclay.cea.fr}
\affiliation{
$^{\dagger}$ E.S.P.C.I, 10 rue Vauquelin, 75005 Paris.\\
$^{+}$ Service de Physique de l'{\'E}tat Condens{\'e},
Orme des Merisiers,
CEA Saclay, 91191 Gif sur Yvette Cedex, France, and\\
Science \& Finance, Capital Fund Management, 6-8 Bd
Haussmann, 75009 Paris, France.\\
}
\date{\today}

\begin{abstract}
We unveil collective effects induced by imitation and social pressure by
analyzing data from three different sources: 
birth rates, sales of cell phones and the drop of applause in concert halls.  We interpret
our results within the framework of the 
Random Field Ising Model, which is a threshold model for collective
decisions 
accounting both for agent heterogeneity and social imitation. Changes of
opinion can occur either abruptly or 
continuously, depending on the importance of herding effects. The main
prediction 
of the model is a scaling relation between the height $h$ of the speed
of variation peak and its width $w$ of the form 
$h \sim w^{-\kappa}$, with $\kappa = 2/3$ for well connected
populations. Our three sets of data are compatible 
with such a prediction, with $\kappa \approx 0.62$ for birth rates,
$\kappa \approx 0.71$ for cell phones 
and $\kappa \approx 0.64$ for clapping. In this last case, we in fact
observe that some clapping 
samples end discontinuously ($w=0$), as predicted by the model for
strong enough imitation.
\end{abstract}

\pacs{Valid PACS appear here}% PACS, the Physics and Astronomy

\maketitle

\section{Introduction}
Traditional economics treat the aggregate behaviour of a whole
population through a ``representative 
agent'' approach, where the heterogeneous preferences of individual
agents are replaced by an average 
preference curve, which determines, for example, the dependence of the
demand on the price of a certain product.
This approach considers that agents determine their action in isolation,
with no reference whatsoever to
the decision of their fellow agents; interactions between agents
are totally neglected. The representative
agent idea has been fiercely criticized by some authors (see e.g.
\cite{Kirman}). Models where the interaction 
between agents is 
explicitly taken into account, traditional in physics, only begin to be
systematically explored in economics and sociology. 
The need to account for interactions stems from the fact that imitation
and social pressure effects are
obviously responsible for the appearance of trends, fashions and bubbles
that would be difficult to understand 
if agents were really insensitive to the behaviour of others. From a
theoretical point of view, as will be clear below, 
interactions lead to an aggregate behaviour that may be completely
different from that implied by a representative
agent approach. Catastrophic events (such as crashes, or sudden opinion
shifts) can occur at the macro level, induced 
by imitation, whereas the behaviour of independent agents would be
perfectly smooth \cite{Granovetter}. 

Imitation is deeply rooted in living species as a strategy for survival:
young children learn (for example languages) 
by imitation. Herds and flocks flee away from danger or keep the correct
destination direction by 
following the motion of their neighbours. Humans are influenced by their
congeners both at a primitive level (fear 
of being excluded from the group) and at a rational level (others may
possess some useful information, revealed by
their very actions). Collective effects induced by imitation can be
beneficial for a society as a whole, but can also 
be detrimental and lead to major catastrophes when imitation cascades
are based on unreliable information or 
dangerous ideas, and when social
pressure supersedes rational thinking. Understanding these collective
effects is therefore of primary importance; 
they may undermine the stability of democracies (and have done so in the
past), lead to crowd panic, financial 
crashes, economic crisis, etc. At a smaller scale, these effects
determine how new products or new technologies 
penetrate a market; strong social imitation can be the key to the
success of a brand, a book or a movie. In any case, 
social imitation often leads to distortion and exaggeration, i.e. a decoupling
between the cause and the effect, which in turn generates inequalities and 
condensation phenomena.

Despite their importance, stressed long ago by Keynes, quantitative
models of imitation effects have only been explored 
in a recent past. One 
branch of models, concerned with information cascades, was initiated by
the paper of Bikhshandani, Hirshleifer and
Welsh in 1993 \cite{BHW} (see \cite{Chamley} for a recent review and
more elaborate developments). Another direction of 
research, where the interaction between 
agents is taken into account explicitly in binary choice situations,
was pioneered by F\"ollmer in 1974 \cite{Follmer}, followed by 
Orl\'ean \cite{Orlean} and others, and, in a sociological context, by
Granovetter in 1978 \cite{Granovetter}, and pursued further in \cite{Granovetter2,Granovetter3}. 
The idea, clearly expressed in Brock and Durlauf \cite{Brock}, is that {\it the
utility or payoff an individual receives depends 
directly on the choices of others in that individual's reference group,
as opposed to the sort of dependence 
which occurs (only) through the intermediation of markets} (see \cite{f1}).  
This category of models have in fact a long history in
physics, where interaction is at the root of spectacular collective
effects in condensed matter, such as ferromagnetism, 
superconductivity, etc. It is therefore not surprising that the
understanding of fashion, booms and crashes, opinion
shifts and the behaviour of crowds or flocks, has attracted considerable
interest in the physics community in the recent 
years (see \cite{Galam} for early insights, and 
\cite{Vicsek,Stauffer,MG,CB,Donangelo,Bettancourt,Chate,HH,Weisbuch,Sornette}
for a short selection).
One particular model, that appears to be particularly interesting and
generic, is the so-called `Random Field Ising 
Model' ({\sc rfim}), which has been successfully proposed to model
hysteresis loops in random 
magnets \cite{Sethna} and a variety of other physical situations. The
hysteresis loop problem is an example of
a collective dynamics of flips (the individual magnetic spins) under the
influence of a slowly evolving external solicitation, but the model can
easily be transposed 
to represent a binary decision situation under social pressure \cite{Galam},
influenced by some global information (such as the price of a product) 
or by {\it zeitgeist}. This transposition was recently discussed in 
several socio-economics context in 
\cite{QF,Book,Nadal}, see also \cite{Rava}. The model has a rich
phenomenology, which we will recall below; 
in particular, discontinuities appear in aggregate quantities when
imitation effects exceed a certain threshold, even 
if the external solicitation varies smoothly with time. Below this
threshold, the behaviour of ``demand'', or of 
the average opinion, is smooth, but the natural trend can be
substantially amplified and accelerated
by peer pressure. The aim of this paper is to explore some situations
where the model should apply, and test qualitative
and quantitative predictions against empirical data. We have gathered
data concerning (a) the drop of birth rates in
European countries in the second half of the XXth century, (b) the
increase of cell phones in Europe in the 90's, (c)
the way clapping dwindles out at the end of music concerts (see \cite{f2})
and (d) crime statistics in
different US states in the period 1960-2000, but this
last data set did not show exploitable idiosyncratic variations. In the
three first cases, we find that our data fits well the
picture suggested by the model, and that noticeable collective effects
can indeed be detected. By analyzing quantitatively 
the shape of the signals, we find that social pressure effects are
distinctly stronger in some countries, or for some
audiences, leading to more abrupt variations. We find a power-law
relation between the maximum slope of the signal 
and the temporal window over which the evolution takes place, in
surprisingly good agreement with the `mean-field' version
of the model. The case of applause is interesting because it is very
close to being a controlled experiment. In that 
case, we observe both continuous and abrupt endings, as predicted by the
model. 

\section{A model for opinion shifts: the RFIM}

We will assume that each agent $i$ is confronted to a binary choice, the
outcome of which being noted $S_i = \pm 1$.
This binary choice can be to vote yes or no in a referendum, it can be to
buy or not to buy a certain good,
to have or not to have children, to clap or to stop clapping, etc. (the
three last examples are indeed studied in the
following sections). We assume that the decision of agent $i$ depends on
three distinct factors \cite{QF,Book,Nadal}: 
\begin{itemize}
\item (i) his personal 
opinion, propensity or utility, measured by a real variable $\phi_i \in
]-\infty,+\infty[$ which we take to be
time independent. Large positive $\phi$'s means a strong a priori
tendency to decide $S=+1$, and large 
negative $\phi$'s a strong bias towards $S=-1$.  
\item (ii) public information, affecting all agents equally, such as
objective informations on the scope of the 
vote, the price of the product agents want to buy, the possibility of
birth control, the advance of technology, etc.
The influence of this time dependent common factor, or polarization
field, will be called $F(t)$, again a real 
variable in $]-\infty,+\infty[$.
\item (iii) social pressure or imitation effects; each agent $i$ is
influenced by the decision made by a certain 
number of other agents $j$ in his ``neighbourhood'', ${\cal V}_i$. The
influence of $j$ on $i$ is taken to be 
$J_{ij} S_j$, that adds to $\phi_i$ and $F$. If $J_{ij} > 0$, the
decision of agent $j$ to buy (say) reinforces the 
attractiveness of the product for agent $i$, who is now more likely to
buy. This reinforcing effect can obviously 
lead to an unstable feedback loop, as discussed below. If on the
contrary $J_{ij} < 0$, the action of agent $j$ 
deters agent $i$ from making the same choice. This ``anti-conformist"
tendency, although rarer in human nature, can 
sometimes exist and be relevant, but this will not be pursued further
here (but see \cite{Granovetter2} for a discussion of this point).
\end{itemize}
In summary, the overall incentive of agent $i$ is $\phi_i + F(t) +
\sum_{j \in {\cal V}_i} J_{ij} S_j$, and the rule 
we choose for the decision of agent $i$ at time $t$ is simply:
\be\label{rfim}
S_i(t) = \mbox{sign}\left[\phi_i + F(t) + \sum_{j \in {\cal V}_i} J_{ij}
S_j(t-1)\right],
\ee
meaning that the decision to ``buy'' is reached when the incentive
reaches a certain {\it threshold value} \cite{Granovetter,Granovetter2}, 
chosen here to
be zero (any other $i$ dependent value could have been chosen, since
this simply amounts to shifting the idiosyncratic 
field $\phi_i$). 
If all $J_{ij} > 0$, the above model is known in physics as the Random
Field Ising model at zero temperature, and has 
been intensely studied in the last decades (see \cite{Sethna} for a
review). In physics, natural networks of 
connections are $d$-dimensional regular lattices, but other topologies,
such as the fully connected case or random
(Erdos-Renyi) graphs have been studied as well. The qualitative
phenomenology does not depend much on the chosen 
topology, nor on the distribution of idiosyncratic fields, although
quantitative details might be sensitive to these.

Let us study first the case where social pressure is absent, i.e.
$J_{ij} \equiv 0, \, \forall (i,j)$. Call $R(\phi)$
the cumulative distribution of $\phi_i$, i.e. the probability that
$\phi_i \leq \phi$. The aggregate demand, or average opinion
${\cal O}$, is defined as: ${\cal O} = N^{-1} \sum_i S_i$, where $N$ is
the total number of agents that we will assume to be 
very large. For a given polarization field $F$, one easily finds:
\be 
{\cal O}_0 = - R(-F) + (1 - R(-F)) = 1 - 2 R(-F).
\ee
(The subscript $0$ means that $J = 0$ here).
As $F$ increases slowly from $-\infty$ to $+\infty$, the average opinion
evolves from $-1$ to $+1$ in a way that mirrors 
exactly the distribution of a priori opinions in the population. For a
generic distribution of $\phi_i$'s (for example,
Gaussian), the opinion evolves smoothly as the polarization field is
increased, as shown in Fig. 1. If one interprets 
$F$ as (minus) the price $P$ of a product, the total demand curve ${\cal
D}$ as a function of the price is:
\be
{\cal D} = N R(P),
\ee
that only reflects individual preferences. 

The situation can change
drastically when imitation is introduced. The
simplest case is when the coupling between agents is ``mean-field'',
i.e. $J_{ij} \equiv 1/N, \, \forall (i,j)$ (see
\cite{Sethna,Nadal} and, in a slightly different setting,
\cite{Granovetter,Granovetter2}). This does not mean that each agent consults all the
other ones before making his mind, but
rather than the average opinion, or total demand, becomes public
information, and influences the behaviour of each
individual agent. This is in fact a very realistic assumption: for
example, the total sales of a book, or number of
viewers of a movie, is certainly an important piece of information for
consumers. It is also known that 
the evolution of the public opinion on a certain topic is affected by
polls, i.e. by a proxy of the average 
opinion. Finally, in the case of financial markets, the price change itself can
be seen, on a coarse-grained time 
scale, as an indicator of the aggregate demand (although the detailed
relationship between the two might 
be quite subtle, see \cite{marketimpact}). This global feedback effect
simply shifts $F$ to $F+J {\cal O}$, leading to a self-consistent
equation:
\be \label{MF} 
{\cal O} = 1 - 2 R(-F - J {\cal O}).
\ee
If imitation is weak enough, one can expand the right hand side in
powers of $J$, leading to first order to:
\be
{\cal O} \approx \frac{{\cal O}_0}{1 - 2 p(F) J},
\ee
where $p(F)={\rm d}R/{\rm d}F$ is the probability density of idiosyncratic fields. 
This equation shows that around the point where the slope of ${\cal
O}_0$ vs. $H$ is maximum, i.e. around  the 
maximum of $p(F)$, the speed of variation of opinion changes is also
maximally amplified -- imitation leads to
exaggeration. As imitation becomes 
stronger, the maximum slope of ${\cal O}$ vs. $F$ increases and finally
{\it diverges} for a critical value $J=J_c$
given by $J_c=A \sigma$, where $\sigma$ is the width of the distribution
$p(\phi)$, and $A$ a numerical constant
that depends on the topology of the graph and on the detailed shape of
$p(\phi)$. This result shows, as expected, 
a diverse population (large $\sigma$) is less prone to collective frenzy
than a more homogeneous one. Above $J_c$, the 
self-consistent equation has, for a range of $F$, three solutions for
$\cal O$, one of which being unstable (see Fig. 1).
This means that as $F$ is increased from $-\infty$, the average opinion
will first follow the lower branch until it jumps 
{\it discontinuously} to the upper branch, for a certain threshold field
$F_c(J)$ (and symmetrically on the way back, 
at $-F_c(J)$, as the field is decreased). This discontinuity is very
interesting from a general point of view: it means that even 
when the external solicitation is slowly and smoothly varied (i.e.
without any information shocks), populations 
as a whole can exhibit sudden, apparently irrational, opinion swings (on
this point, see \cite{Granovetter} for 
an early discussion -- in particular his Fig. 2). In an economic context, it
means that the demand for a product can vary discontinuously from low to high 
as the price is decreased \cite{Granovetter2,Nadal}.

\begin{figure}
\begin{center}
\psfig{file=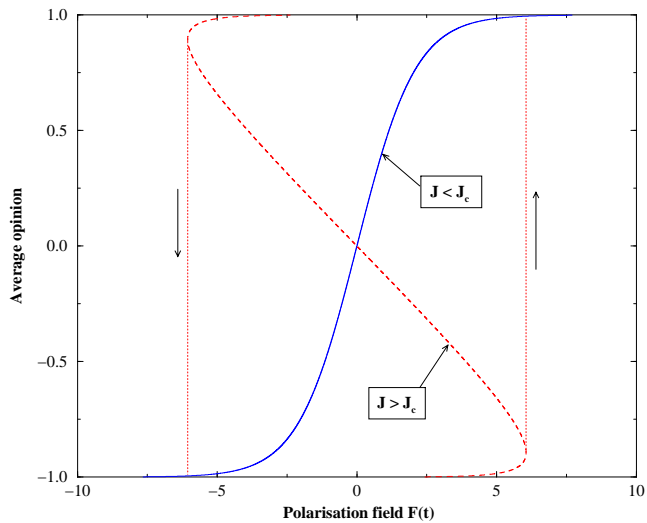,width=7cm,angle=270} 
\end{center}
\caption{Average opinion ${\cal O}$ as a function of the external
solicitation field $F(t)$. For an imitation 
parameter $J < J_c$, the curve is smooth; whereas for $J > J_c$ an
hysteresis effect appears: the average opinion
remains small for an anomalously large value of $F(t)$ before suddenly
jumping to the upper branch. The
amplitude of the jump increases as $(J-J_c)^{\beta}$ for $J - J_c$ small
\cite{Sethna}.}
\label{Fig1}
\end{figure}

The vicinity of the critical point $J \stackrel{<}{\sim} J_c$ reveals
interesting critical properties, to a large extent
independent of the detailed form of $p(F)$ or $R(F)$. Noting
$\varepsilon = J_c - J$ the distance from criticality,
one finds that the opinion slope ${\cal S}={\rm d}{\cal O}/{\rm d}F$ takes a scaling
form:
\be \label{scaling}
{\cal S} = \frac{1}{\varepsilon}\,  {\cal G}
\left(\frac{F-F_c(J)}{\varepsilon^{3/2}}\right),
\ee
where the function ${\cal G}(x)$ can be computed explicitly
\cite{Sethna}, and is universal. Its shape is plotted 
in Fig. 2; one finds that ${\cal G}(0)$ is a finite constant and ${\cal
G}(x \to \infty) \sim x^{-2/3}$.
Eq. (\ref{scaling}) means that the slope, as a function of $F$, peaks at
a maximum of order $\varepsilon^{-1}$ and 
remains large on a small window of field of order $w \sim
\varepsilon^{3/2}$. In other words, the height $h$ of the peak
behaves as $w^{-\kappa}$ with $\kappa=2/3$, instead of $\kappa=1$ for a
regular evolution, without any collective effects ($J \equiv 0$). 
These results are expected to be rather robust: not only the
distribution of idiosyncratic fields is (within a broad class)
irrelevant, but also the detailed topology of the graph. 
Only if the graph is a regular lattice of dimensions less than $d=6$
will the shape of the scaling function and
the value of the exponent $\kappa$ be affected \cite{Sethna}. Beyond $J_c$, the curve
${\cal O}(H)$ becomes discontinuous; the 
amplitude of the jump is found to increase as $(J-J_c)^{\beta}$ for $J -
J_c$ small, with $\beta=1/2$ in 
mean-field situations such as the one described above \cite{Sethna}.

If now one `zooms' on fine details (on scale $1/N$) of the curve ${\cal
O}(F)$ in the vicinity of $J=J_c$, one finds that the 
evolution of $\cal O$ is actually resolved in a succession of `avalanches' of
different sizes $s$, where $s$ agents simultaneously
change their opinion. Interestingly, the distribution of avalanche sizes
also takes a scaling form which has 
attracted enormous interest in the context of random magnets
\cite{Sethna}, and offers an enticing microscopic picture of 
how large opinion swings actually develop in a population: as a
succession of events, most of which small, but some 
involving an extremely large number of individuals. More precisely, at
the critical point $J=J_c$, 
the distribution of avalanche sizes as $F$ is swept through $F_c$ decays
for large $s$ as $s^{-\mu}$ with $\mu=9/4$ \cite{Sethna}.

In summary, the transposition of the {\sc rfim} to opinion shifts
predicts that the average opinion evolves,
as a function of the global solicitation, very differently if imitation
effects are weak (in which case the 
evolution is smooth) or if imitation exceeds a certain threshold, in
which case aggregate quantities are 
discontinuous and catastrophic avalanches may be triggered. More
quantitatively, the model predicts that in generic
situations, the evolution slope should peak more and more as the
critical point is approached: its height $h$ increases
while its width $w$ decreases, the two being related by $h \sim
w^{-2/3}$. We will now show that these predictions 
seem indeed to be relevant in the three situations that we have
investigated: evolution of birth rate, of cell (mobile)
phones and of clapping activity. 
The idea is that different countries, or different crowds,
are characterized by different values of the heterogeneity parameter
$\sigma$ and/or of the imitation parameter $J$. Therefore, the distance
from the critical point $\varepsilon$ will vary across the set of
available countries or audiences, leading to peaks of different height
and width.
[A fourth set of data, concerning crime
statistics in the US, was inconclusive, since the
height $h$ and width $w$ were not found to vary significantly from state
to state.] 

\begin{figure}
\begin{center}
\psfig{file=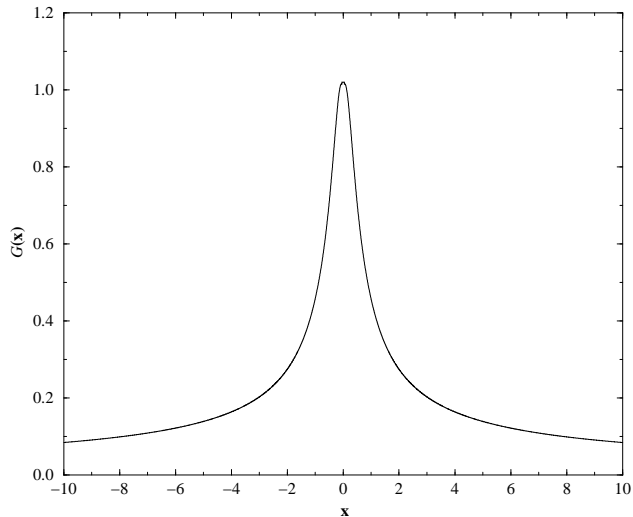,width=7cm,angle=270} 
\end{center}
\caption{Shape of the scaling function of the peak, ${\cal G}(x)$,
defined by Eq. (\ref{scaling}).}
\label{Fig2}
\end{figure}

\section{Birth rate in Europe 1950-2000}

The decision to have children has been profoundly affected by an easier
access to birth control, higher education, 
the loss of influence of religions, etc, which play the role of a slowly
evolving global solicitation field $F(t)$. 
There might also be a significant social pressure effect: the image 
of women in society, the social prestige of having a career rather than
a family life, may have a strong deterring 
influence of the choice of having, or not having children. Therefore, we
believe that the three sources of influence
appearing in Eq. (\ref{rfim}) above should be important to understand
the detailed evolution of birth rate in the 
second half of last century.

\begin{figure}
\begin{center}
\psfig{file=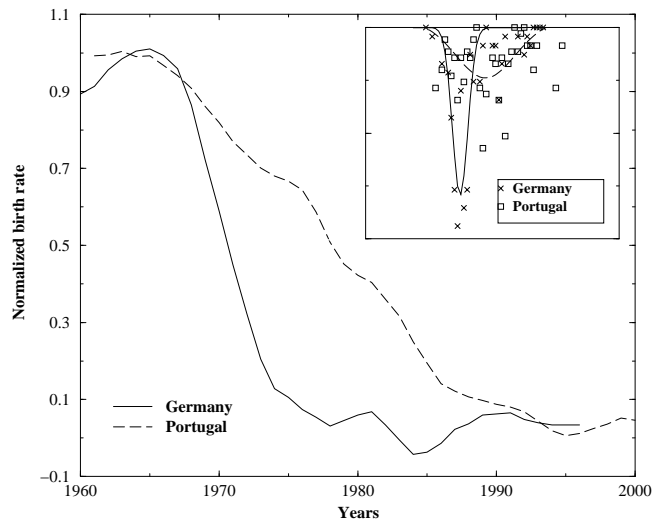,width=7cm,angle=270} 
\end{center}
\caption{Normalized fecundity index as a function of time for Germany
and Portugal (3 year average). Other countries 
are intermediate in terms of the sharpness of the crossover.
Inset: Yearly change of the fecundity index and 
Gaussian fits, allowing one to extract both the height $h$ and width $w$
of the peaks.}
\label{Fig3}
\end{figure}

We have downloaded from Eurostat (europa.eu.int/comm/eurostat/) the
birth rate in several representative 
European countries (Belgium, France, Germany,
Grece, Italy, 
Netherlands, Poland, Portugal, Spain, Sweden, Switzerland, UK). The
reported birth rate is the average fecundity index
in a given year, from 1950 to 2000, except for Germany where we only
keep data before re-unification. In all countries, except Sweden,
this index has steadily decreased over the years. In Sweden, the birth
rate has, after 
having decayed like in other countries, increased back in the late
eighties, probably due to governmental incentives. 
In most countries except Grece and Portugal, the slope of the decay
reached a maximum around 1970. We have 
treated the data as follows (an identical treatment will also be applied
in the following sections). We 
first re-scale the data such that the total range of variation is
constant, independent of the country, such as
to remove any idiosyncratic, cultural effects on the average number of
children per family. We then define the slope
${\cal S}(t)$ by taking the discrete derivative of the curve, defining
the yearly change of (rescaled) birth rate. 
This leads to a rather noisy curve, but 
that has a distinctive peak as a function of time, see Fig. 3. We then
fit ${\cal S}(t)$ using a Gaussian shape for 
the peak, 
plus a constant background (found to be small compared to the maximum
height of the peak). Since the data is noisy, 
the Gaussian shape is only a convenient way to extract
the height of the peak $h$, its width $w$ and the location of the peak
$t^*$. Interestingly, we find a rather 
large variability of $h$ and $w$ across countries, much beyond the
uncertainty of the fitting parameters $h$ and $w$.
Some countries, like West Germany, display a high and narrow peak while
others, like Portugal, have
a much broader peak, with a more modest height (see Fig. 3). Now,
plotting $\ln h$ vs. $\ln w$ across
different countries, we find (see Fig. 4) that the data clusters
reasonably well around a straight line 
of slope $-2/3$, 
as predicted by the fully connected {\sc rfim} detailed in the previous
paragraph, and in 
any case significantly smaller than the trivial value $-1$. Remember
that the usefulness of that
prediction is that it does not require to specify the distribution of
idiosyncratic fields $\phi$. 
The best regression through the data in fact gives a slope of $-0.71 \pm
0.11$. This suggests that the accelerated birth drop in some countries 
is indeed induced by a social pressure effect, 
rather than being a mere dependence of diversity
(measured by variance $\sigma^2$ of $p(\phi)$) on the country, since in that case 
one would trivially observe $h \sim 1/w$. Note also
that the significant change of height and width 
across different countries makes rather unplausible that the drop in the
birth rate is solely due to a sudden
external cause, like the availability of birth control pills. Although
this availability certainly triggered
the phenomenon, its amplification, according to our analysis, appears to
be compatible with imitation effects, 
exactly as described by the {\sc rfim} with a slowly evolving external
drive $F(t)$.

\begin{figure}
\begin{center}
\psfig{file=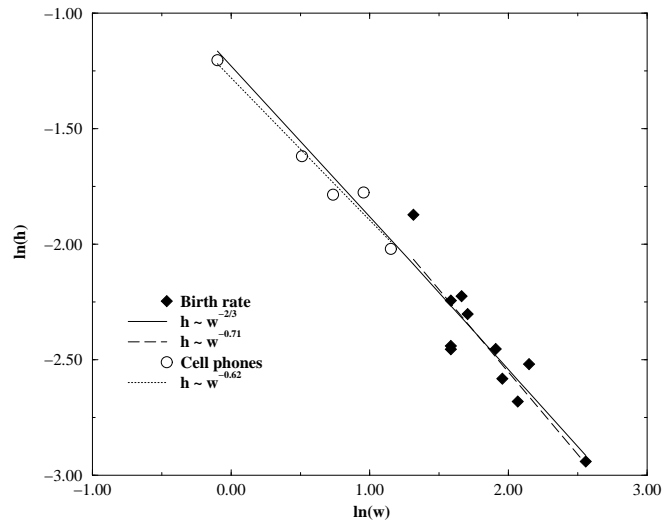,width=7cm,angle=270} 
\end{center}
\caption{Height of the peak $h$, vs. width of the peak $w$, in a log-log
scale, both for birth rates and for cell
phones. The mean field prediction $h \sim w^{-2/3}$ is shown for
comparison. A typical relative error 
of $20\%$ on the fitted values of $h$ or $w$ translates into vertical and
horizontal error bars of $0.2$, 
comparable with the erratic spread of the points. The heights
corresponding to 
cell phone data has been divided by a factor $1.7$ to match the birth
rate data (the absolute height of the peak is in fact not universal).
The width $w$ is however not rescaled, and shows that the explosion of
cell phones is, as expected, 
faster that the collapse of the birth rate.}
\label{Fig4}
\end{figure}

\section{Number of cell phones in Europe, 1994-2003}

The evolution of the number of cell phones in the last decade is an
interesting case study. The soaring number of 
cell phones obviously followed better technology, lower prices, etc.,
encapsulated in our global solicitation
field $F(t)$, which we again assume to be smoothly evolving with time
(no technology shocks). But it is also clear 
that social effects must have been important, not only as in usual
fashion 
phenomena where not possessing the trendy object gives rise to a feeling
of inferiority, but also because the usefulness
of cell phones is objectively increased if more people can use one. 

We have studied the monthly evolution of the total number of cell phones in use (all
providers included) in five different European 
countries (Germany, UK, France, 
Italy, Spain) in the period 1994-2003. We obtained the data from the
{\sc art} 
(``Autorit\'e de R\'egulation des T\'el\'ecommunications'' in Paris). During that
period, the number of cell phones increased 
by a factor $\sim 20$ or more. We apply to the data the same treatment 
as in the previous section: normalization, discrete difference and a
Gaussian fit. The normalized data is plotted 
in Fig. 5. From the Gaussian fit of the peak, we again extract a height
$h$, width $w$ and location $t^*$. In this case,
we find that all curves peak at $t^*=2000.35 \pm 0.25$. As for birth
rates, we find a rather large spread in the values of $h$ and $w$, 
with again Germany standing out as the most ``collective'' country. What
we find quite remarkable is that the five
points fall quite nicely on the same curve as for birth rates, although
the time scale is obviously shorter for 
the spreading of cell phones than for the decay of birth rates (see Fig.
4). Notice however that the birth rate collapse 
in Germany is almost as fast as the rise of cell phones in Italy! The
slope of the best regression over 
the five points is $-0.62\pm 0.07$, clearly compatible with a
slope $-2/3$ (see Fig. 4). This result rather strongly suggests 
some social amplification of the trend, compatible with the self-referential 
mechanism proposed here.

Note that although quite sharp, the above curves all seem to lie in the
smooth regime $J < J_c$ of the {\sc rfim}. 
On the other hand, a true runaway behaviour, predicted for $J > J_c$,
cannot occur in the case of cell phones 
because the rate at which these can be commercialized is finite; this
effect would regularize the discontinuity and
give it a finite width related to the maximum sale capacity of phone
providers. This regularisation effect will indeed
be observed in our last example, to which we now turn. 

\begin{figure}
\begin{center}
\psfig{file=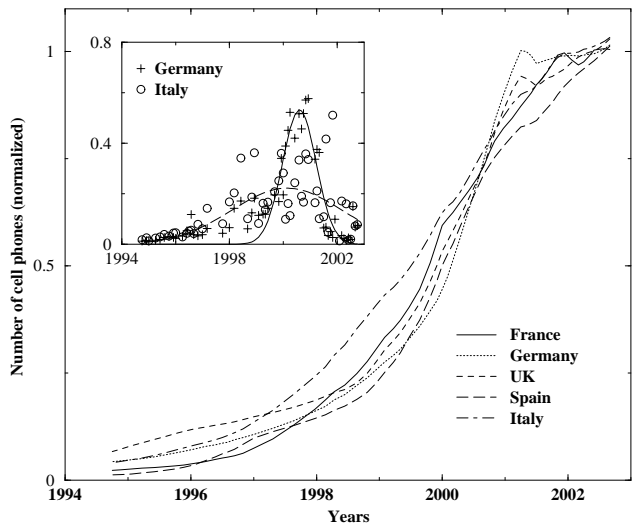,width=7cm,angle=270} 
\end{center}
\caption{Evolution of the total number of cell phones in use (all
providers included) in various European countries (3 month average).
Inset: Monthly change 
for Germany and Italy, allowing one to extract both the height $h$
and width $w$ of the peaks.}
\label{Fig5}
\end{figure}

\section{Persistent and interrupted clapping}

After a concert, or a theater performance, the public usually claps for a
period of time that reflects its satisfaction or
enthusiasm. A collective effect which is well documented but that we will
not be concerned with here is the possible {\it synchronization} of claps \cite{Clapping} 
(except insofar as these synchronization effects already suggest the importance 
of interactions between individuals). We rather want to focus
on the way the clapping dies out. The point
here is that if people have a different degree of enthusiasm for the
performance, the time after which they will 
stop clapping varies from one person to the next. More precisely, the
idiosyncratic field $\phi_i$ defined above is,
in the present case, the {\it a priori} amount of time $t_i$ a given
individual would carry on clapping if isolated from 
others. But of course, we all hear what others are doing, and are
clearly influenced by the level of clapping of the
public as a whole. Many people would hate being the last individual to
clap in a large concert hall. Therefore, the
propensity to clap adapts to the overall clapping intensity, exactly as
in the {\sc rfim}. In the present case, the
external field $F(t)$ is simply (minus) the time elapsed since the start
of applause: as time passes, the necessity
to clap declines. Therefore, in the absence of interaction between
individuals, the state $n_i=1,0$ of agent $i$
is given by: 
\be
n_i = \Theta\left(t_i - t\right),
\ee
where $\Theta$ is the Heaviside function, $\Theta(x>0)=1$, $\Theta(x <
0)=0$. The overall sound level $\cal I$ 
is proportional to the number of clapping people $\sum_i n_i$ (we
neglect here the individual fluctuations in clapping intensity); 
the model we propose is simply that the {\it a priori} clapping time
$t_i$ is shifted by a quantity proportional 
to $\cal I$: $t_i \longrightarrow t_i + J {\cal I}$; therefore the total
clapping intensity at time $t$ is 
given by:
\be
{\cal I} = 1 - R(t - J {\cal I}),
\ee
where we have just repeated, for the sake of clarity, the general
argument given in section 2, adapted to the present case.
Therefore, we expect two distinct types of applause: ``persistent''
clapping that slowly die out with time, 
as more and more people progressively decide to stop clapping ($J <
J_c$), and ``interrupted'' clapping, which end 
abruptly, because people pay acute
attention to the global clapping level and stop clapping as soon as they
hear that the noise level starts decreasing
($J > J_c$). 

\begin{figure}
\begin{center}
\psfig{file=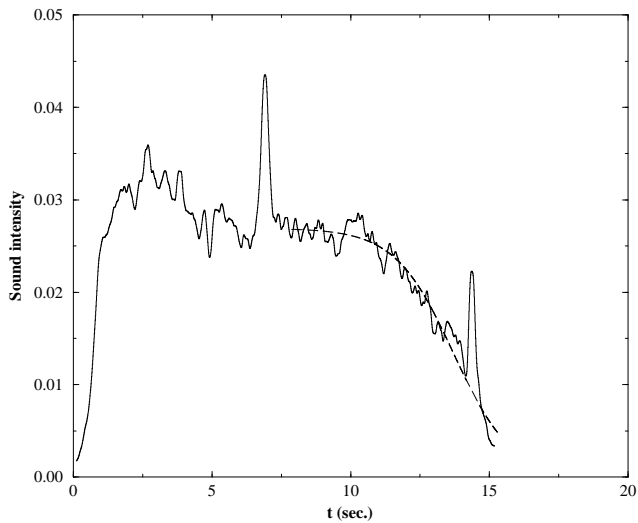,width=7cm,angle=270} 
\end{center}
\caption{Typical time series of sound intensity as a function of time,
during applause. One sees the initial rise, 
a relatively constant plateau phase, and the final phase where clapping
is tapering off. Notice a few spurious spikes, 
corresponding to occasional `bravos' or other shouts. The data was filtered with
a Gaussian window of width $0.225$ sec.}
\label{Fig6}
\end{figure}

We have analyzed data coming from music concerts recorded in the very
same concert studio, ``Studio 104'' 
of Maison de la Radio (Paris, France). The concerts took place 
during two special events called ``Couleurs Francophones'', that featured a total 
of 11 performances of french speaking artists both in March 2003 and March 2004. 
The data therefore concerns
two audiences of ca. 1000 people, each reacting to 11 different performances of
various quality, leading to different degree and variability of enthusiasm of the crowd.  
Each recording is stereo with two channels; we were 
able to obtain the raw data, not post-processed by sound engineering. Most of the available data
is actually post-processed, for example, the clapping phase is artificially cut-off.

A typical recording of the clapping is shown in Fig. 6. After an initial fast rise, 
one sees a relatively
constant level of clapping, followed in this case 
by a rather
smooth decay of the sound intensity. Note however the presence of a few peaks,
corresponding to occasional shouts or other 
perturbations. At the qualitative level, we indeed find two different
categories of clappings, as illustrated in Fig. 7, 
where we show a slow persistent clapping, that takes over 10 seconds to 
smoothly die out, and a two fast ones. For the
fastest case, the width of the stopping period 
is in fact given by the width of the filter that we used, which is
comparable to the acoustic decay time of 
the room ($\sim 1$ sec). This shows that within the measurement
accuracy, the stopping period was instantaneous, 
corresponding in our model to strong imitation effects, $J > J_c$. The
possibility of observing such discontinuous
events makes the clapping data extremely interesting: as mentioned
above, both birth rate data or cell phone data 
seem to be in the continuous regime ($J < J_c$).

\begin{figure}
\begin{center}
\psfig{file=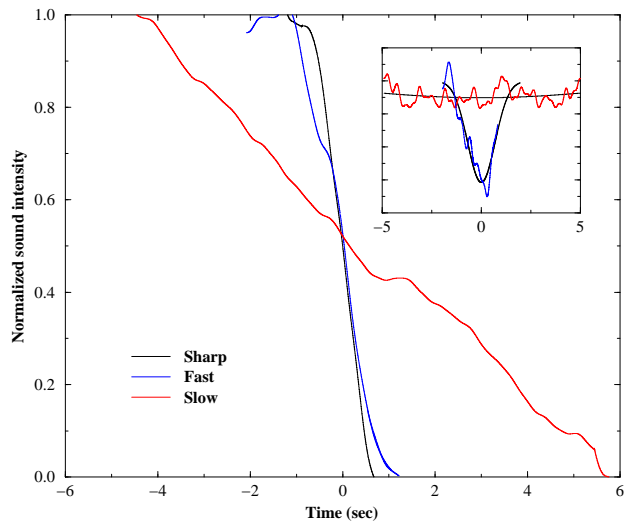,width=7cm,angle=270} 
\end{center}
\caption{Terminal stages of clapping corresponding to three
characteristic recordings: one of them is a slow 
decay of applause (over 10 seconds), corresponding to a very heterogeneous audience. The
two other ones are fast events (on the order of one second), one
of them can even be classified as instantaneous since its width cannot
be resolved (i.e. it is thinner than the Sabine
reverberation time of the room ($\approx 1.8$ sec)).}
\label{Fig7}
\end{figure}

More quantitatively, we have again extracted from this data set heights
$h$ and widths $w$ of the peaks in the 
sound intensity change, following exactly the same procedure as above.
In this case, we have removed a few 
recordings that are extremely noisy (many spurious peaks, very different
left and right channels) or atypical 
(one recording shows a two stage decay, which can happen in the case of
a very strongly heterogeneous audience: half
of the audience may have stopped clapping, while the other half carries 
on). This leaves us with 17 different samples out of the 22 at our disposal. 
The result is plotted in Fig. 8, in a log-log scale, where we do not
report the 2 points corresponding to `abrupt' endings ($< 1$ sec). 
Rather remarkably, we find the very same quantitative pattern as in Fig.
4: the heights of the peaks are related to
their width as $h \sim w^{-0.64}$, again very close to the mean-field
prediction of the {\sc rfim}. We find this 
result quite striking, since this example is very close to being a
controlled experiment, that clearly gives 
some credibility to the idea of social amplification trends, in a way
fully compatible with the {\sc rfim} 
framework. We regard the existence of abrupt endings in a crowd of 1000 people, 
predicted by the model but unobserved in the previous two
examples, as a strong support of our contention. However, our
analysis should be redone on a different, larger set 
of applause endings in well controlled situations, to consolidate our
findings.

\begin{figure}
\begin{center}
\psfig{file=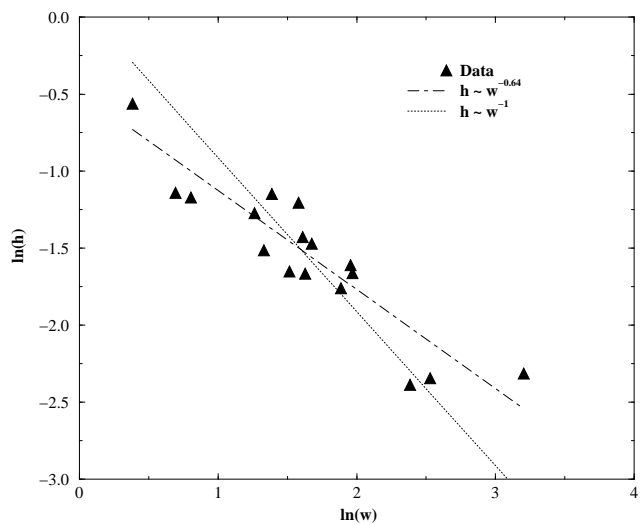,width=7cm,angle=270} 
\end{center}
\caption{Height of the peak $h$, vs. width of the peak $w$, in a log-log
scale, for 17 applause endings. A 
best fit (in log scale) leads to a slope of $-0.64 \pm 0.07$, again very
close to the mean-field {\sc rfim} 
prediction of $-2/3$. Note that the above value of the error bar comes
only from the regression. We have also shown, for comparison, the slope 
corresponding to a `trivial' behaviour, $h \sim 1/w$.}
\label{Fig8}
\end{figure}

\section{Conclusion}

In this work, we have tried to document quantitatively collective
effects induced by imitation and social 
pressure. We have analyzed data from three different sources (birth
rates, sales of cell phones, decay of 
applause) within the framework of the Random Field Ising Model, which
is a threshold model for collective decisions 
accounting both for agent heterogeneity and social imitation. Changes of
opinion, demand or behaviour in this 
model can occur either abruptly or continuously, depending on the
importance of herding effects. The speed of
change generically peaks at a certain time; the main prediction of the
model is a scaling relation between the 
height $h$ of the peak and its width $w$ of the form $h \sim
w^{-\kappa}$, with $\kappa = 2/3$ for densely 
connected populations (mean-field situation). Our three sets of data are
compatible with such a prediction, with 
$\kappa \approx 0.62$ for birth rates, $\kappa \approx 0.71$ for cell
phones and $\kappa \approx 0.64$ for clapping. 
In the last case, we in fact 
observe that some clapping samples end discontinuously ($w=0$), as
predicted by the model for strong enough 
imitation. Since the data is rather noisy, the agreement of each
individual example with the model is perhaps not
particularly impressive. However, we believe that the convergence of
these three rather different situations and
the robustness of the theoretical picture gives some credit to our
conclusions. 

Many other situations could be analyzed according to similar lines: the
invasion of other products and 
technologies (cars, television, Internet access, etc.), or other social
phenomena such as divorces, car accidents
(for example, the death toll on the French roads has sharply decreased
in the past few years after staying 
among Europe's highest for decades), the evolution of opinion polls,
social riots, strikes or upheavals \cite{Granovetter}, residential 
segregation \cite{Granovetter3}, discontinuities in
History, etc., provided of course reliable data is available. The case
of crime statistics is interesting and has been the 
subject of some studies (see \cite{JS}). We have actually downloaded the
statistics of crime in different 
US states in the period 1960-2000. These show a rather sharp rise during the
sixties; unfortunately, all states behave 
more or less identically, and the study performed above could not be
carried through. The conclusion is, perhaps
not unexpectedly, that the US population tends to behave much more
uniformly than in Europe, where cultural
differences are very perceptible (as our results on birth rates and cell
phones confirm).

The present work could be extended in various directions: getting better
data, on a wider set of examples and 
samples, would confirm or disprove the validity of our analysis. For
example, the direct observation of avalanches 
of opinion changes, and the distribution of their size, would be highly
interesting. Also, situations where the coupling is not
global but more local, such as, e.g. large open air concerts, would be
worth investigating since the value of 
$\kappa$ is expected to change, and gets closer to $1$ for
two-dimensional geometries \cite{Sethna}. From a theoretical point
of view, one 
could include in the {\sc rfim} truly dynamical effects, in the sense
that crowds are not only sensitive to the 
aggregate opinion, but also to the speed of change of this opinion \cite{Bettancourt},
which might lead to even stronger instabilities.
A way to include these in the {\sc rfim} framework is to rewrite Eq.
(\ref{MF}) in a differential form, and to
add an extra `speed of change' feedback, in the spirit of \cite{BC}:
\be 
\left[1 - 2J p(-\Phi)\right] \frac{{{\rm d}\cal O}}{{\rm d}t} = 2 R(-\Phi)
 V + 
\int_{-\infty}^t d\tau K(t-\tau) \frac{{{\rm d}\cal O}}{{\rm d}\tau}, \qquad \Phi = F + J{\cal O},
\ee
where $V$ is the rate of change of $F$ and $K(.)$ is a certain memory
kernel. 

An interesting outcome of a deeper understanding of collective effects
would be to be able to disentangle, in the evolution of public opinion, 
a genuine opinion shift from self-referential effects. If this
separation could itself become public information, 
the resulting `holoptical' feedback loop (following the term proposed by
Noubel \cite{Noubel}) could be stabilizing
and prevent, or at least temper, opinion swings, financial crashes or
economic crises. 

\begin{acknowledgments} 
J.P. B thanks A. Kirman, M. Marsili, M. A. Miceli, J.P. Nadal, J.F.
Noubel, A. Orl\'ean, J. Sethna and 
R. da Silveira for very useful conversations on these subjects over the
years. We also thank Edwige Ronci\`ere and Laurent Givernaud, from Radio France, 
for providing the clapping recordings, and M. Granovetter for pointing us to 
refs. \cite{Granovetter2,Granovetter3}.
\end{acknowledgments}


\begin{thebibliography}{99}
\bibitem{f1} The word ``only'' was added by us. The fully connected limit of 
the {\sc rfim} analyzed below however corresponds to the case where the interaction 
is in fact mediated by  global information, such as the price.
\bibitem{f2} These examples were actually proposed in \cite{QF,Book}.
\end{thebibliography}
\end{document}